\documentclass[letter,12pt,english]{article}
\usepackage{graphicx,amsmath,amsfonts}

\newcommand{\beq}{\begin{equation}}
\newcommand{\eeq}{\end{equation}}
\newcommand{\beqr}{\begin{eqnarray}}
\newcommand{\eeqr}{\end{eqnarray}}
\newcommand{\beqrn}{\begin{eqnarray*}}
\newcommand{\eeqrn}{\end{eqnarray*}}
\newcommand{\barr}{\begin{array}}
\newcommand{\earr}{\end{array}}
\newcommand{\bef}{\begin{figure}}
\newcommand{\eef}{\end{figure}}
\newcommand{\bec}{\begin{center}}
\newcommand{\eec}{\end{center}}

\newcommand{\eps}{\epsilon}\newcommand{\om}{\omega}

\begin{document}

\title{\textbf{Frequency Content and Autocorrelation Function of Noisy Periodic Signals}}
\author{D. Villani$^{1}$\footnote{To whom correspondence should be addressed
 (dvillani@alumni.princeton.edu).}
, R. M. Ghigliazza$^{2}$, R. Carmona$^{3,4,5}$ \\ \\
$^1$Hess Energy Trading Company, LLC \\
1185 Avenue of the Americas, New York, NY 10036 \\
$^2$Department of Mechanical and Aerospace Engineering \\
Princeton University, Princeton, NJ 08544 \\
$^3$Department of Operations Research and Financial Engineering \\
Princeton University, Princeton, NJ 08544\\
$^4$Bendheim Center for Finance\\
Princeton University, Princeton, NJ 08544 \\
$^5$Applied and Computational Mathematics Program\\
Princeton University, Princeton, NJ 08544} \maketitle

\begin{abstract}
We extract the frequency content of a noisy signal by use of
Discrete Fourier Transform. Our analysis overcomes the limitations
imposed by incommensurate lattices. After computing the
deterministic component, we show the relevance of the method in
removing spurious autocorrelations from the signal residuals.
Results are presented for a temperature time series.
\end{abstract}

\bigskip \pagebreak

\section{Introduction}
The computation of the autocorrelation function of a noisy signal
usually requires the removal of a deterministic component often
called trend or seasonal component. Motivated by the important
problem of the statistical analysis of temperature data, we
concentrate on the case of noisy periodic signals. For this
particular application, we need to carry out the following steps.
(i) First, we evaluate the frequencies of the embedded periodic
components. (ii) Then, we fully identify the deterministic
component by a variational principle restricted to the class of
functions consistent with the results of part (i). (iii) Finally,
the computation of the autocorrelation function of the residuals
completes the analysis of the signal. We discuss the theoretical
underpinnings of such a method in the special case of a signal
with a single periodic component, and we demonstrate that this
program is often spoiled by the subtle consequences of the
possible incommensurability of the sampling frequency and the
intrinsic frequency of the signal in question. See for example
\cite{Carmona} for a discussion of the sampling theory of
continuous signals. This paper quantifies one form of this
undesirable effect, and proposes a remedy to the resulting
ambiguity in the determination of the intrinsic frequency of a
noisy periodic signal.

\section{Fourier Spectrum}
To establish notation, we begin by giving the explicit formula for
the Discrete Fourier Transform (DFT for short) $X_k$ of a given
vector $x_j$ of length $N$ \cite{Cizek}: \beq X_k = \frac{1}{\sqrt
N}\sum_{j=1}^{N} x_j e^{ 2\pi i \frac{j-1}{N} (k-1)},\qquad
k=1,\cdots,N. \eeq With this definition of the Fourier transform,
it follows that: \beq x_j = \frac{1}{\sqrt N}\sum_{k=1}^{N} X_k
e^{-2\pi i \frac{k-1}{N} \, (j-1)},
\label{eq_def} \eeq because of the orthogonality property \beq
\frac{1}{N} \sum_{j=1}^{N} e^{-2 \pi i
\frac{n-k}{N}j}=\delta_{nk}. \label{eq_orth} \eeq Here and in the
following, $\delta_{nk}$ is the Kronecker delta.
\vskip 2pt
\noindent Let us consider a monochromatic signal: \beq x_j = A
e^{-2 \pi i\frac{m-1}{N}j}, \qquad j=1,\ldots, N,
\label{eq_monochr} \eeq where $m$ is a given positive integer between $1$
and $N$. For each integer $1\le k\le N$ we have:
\beq
\label{eq_FT_monochr} X_k = A \sqrt N \; e^{-2 \pi i
\frac{k-1}{N}} \; \delta_{km}, \eeq and hence: \beq \left|X_k
\right| = \left|A \right| \sqrt N \; \delta_{km}. \eeq In other
words, the spectrum results in all coefficients being zero apart
from a peak for the value of $k$ equal to $m$. From the spectrum
we can compute the frequency of the periodic signal given the
length $N$ (i.e., $\om = \frac{m-1}{N}$). A difficulty arises
when the monochromatic signal is of the form: \beq x_j = A e^{-2
\pi i\frac{m-1+\eps}{N}j}, \qquad j=1,\ldots, N, \qquad
-\frac{1}{2} \le \eps < \frac{1}{2}. \label{eq_monochr_inc} \eeq
In this case, the frequency cannot be expressed as a ratio of the
form $(m-1)/N$ and we say that we are facing an
\emph{incommensurate} lattice problem. Computing the spectrum, we
get: \beq \left|X_k \right| = \left|A \right| \frac{1}{\sqrt N}
\sqrt{\frac {1-\cos{2 \pi \left(m-k + \eps \right)}} {1-\cos{2
\pi\frac{m-k + \eps}{N}}}}.
\label{eq_FT_mnchr_inc} \eeq It is now evident that an
incommensurate frequency produces a spread of the peak (i.e.,
$X_k \neq 0$ for $k \neq m$). This is illustrated in Figure
\ref{fi:surface}.
\bef \bec
\includegraphics[scale=0.54]{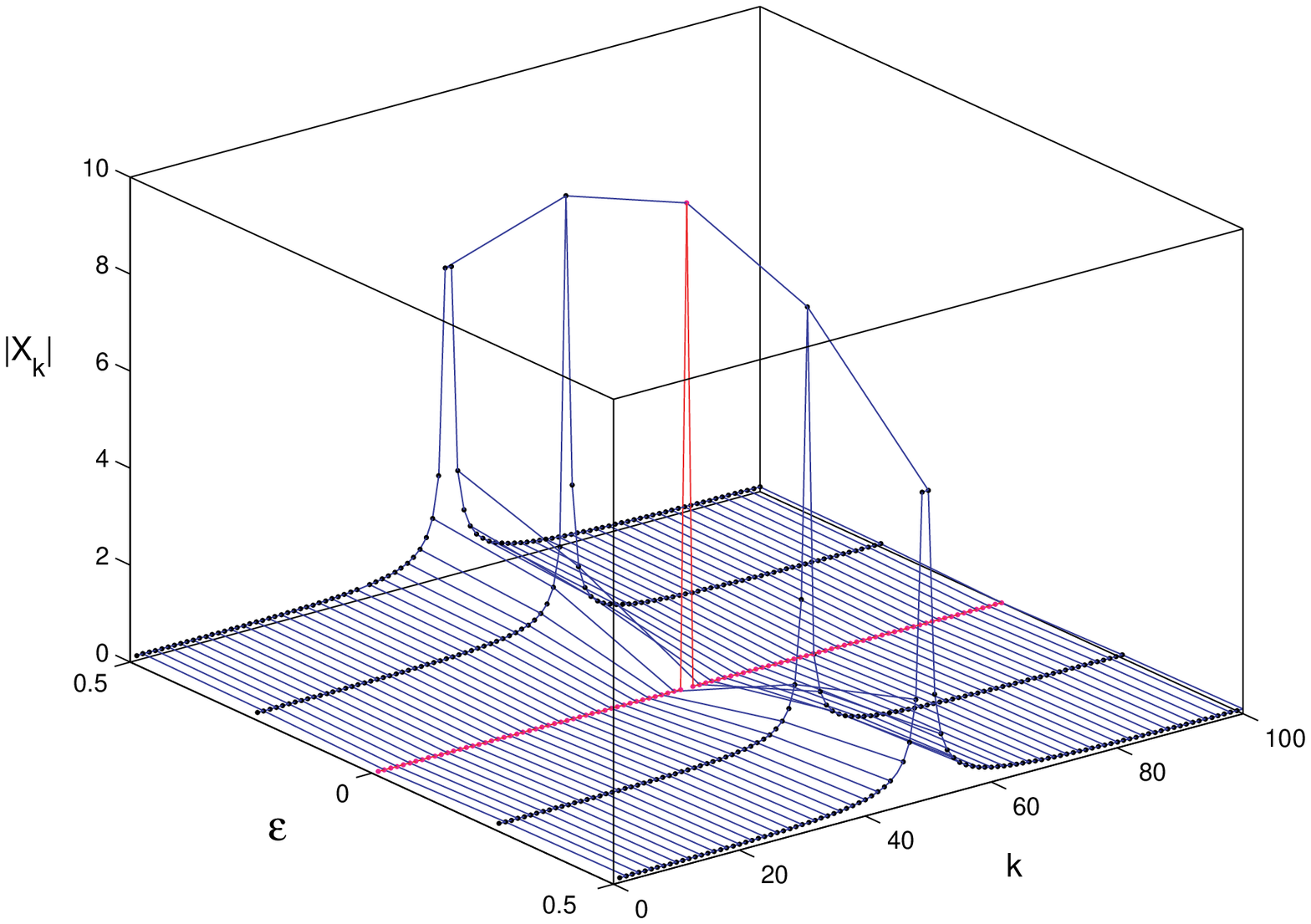}
\caption{The spectrum $|X_k|$ as a function of $k$ and $\epsilon$.
$|A| = 1$, $N=100$ and $m=50$. The solid lines are a guide to the
eye.} \label{fi:surface} \eec \eef
It is worth noting that for $\eps \rightarrow 0$ and $N
\rightarrow \infty$, we obtain \beq \left| X_k \right| \approx
\left|A\right| \sqrt{N} \left| \frac{\sin \pi (m-k)}{\pi (m-k)}
\right| \left\{ 1 + \left[ \pi \cot \pi (m-k) -\frac{1}{m-k}
\right] \; \eps \right\} + O(N^{-3/2}). \label{eq_exp} \eeq We
recognize the zero order term as the Fraunhofer diffraction
pattern for the single slit \cite{BornWo}. This should not be
surprising as the spread of the Fourier spectrum $X_k$ is due to
the interaction of two wavelengths; one for the lattice and the
other for the signal.

The previous analysis shows that the Fourier spectrum is widened
even in the case of a noise free signal. In the case of a noisy
signal the detection of the frequency is complicated by the
spectrum asymmetries introduced by the noise.

\section{The Case of Temperature Data}
We now analyze the average temperature data for the case of
Seattle-Tacoma international Airport from $1/1/1960$ to
$12/31/1999$ (i.e., $14,610$ entries). First, we remove the mean
value of $52.20$ Fahrenheit. Next, we determine the frequency of
the embedded periodic signal by use of the Fourier paradigm. For
any subset of the data, the Fourier spectrum gives a peak over a
noisy background. However, each peak gives different estimates of
the period, all inconsistent with the naive idea that the more
points the better the estimate of the period. In Figure
\ref{fig_om_N}, we show the value of the estimated period
$\Lambda$ as a function of $N$. The $\Lambda$'s were computed
according to the following prescriptions. For each $N \le 14,610$
we selected the first $N$ points of the data set and computed the
period as \beq \Lambda = \frac{N}{k_{\text{max}}-1},
\label{eq_lambda} \eeq where $k_{\text{max}}$ is the index for
which the absolute value of the spectrum is maximum. We always
chose the integer $k_{\text{max}}$ to be in the range below the
Nyquist frequency to avoid aliasing artifacts \cite{Priest}.
\bef \bec
\includegraphics[scale=0.54]{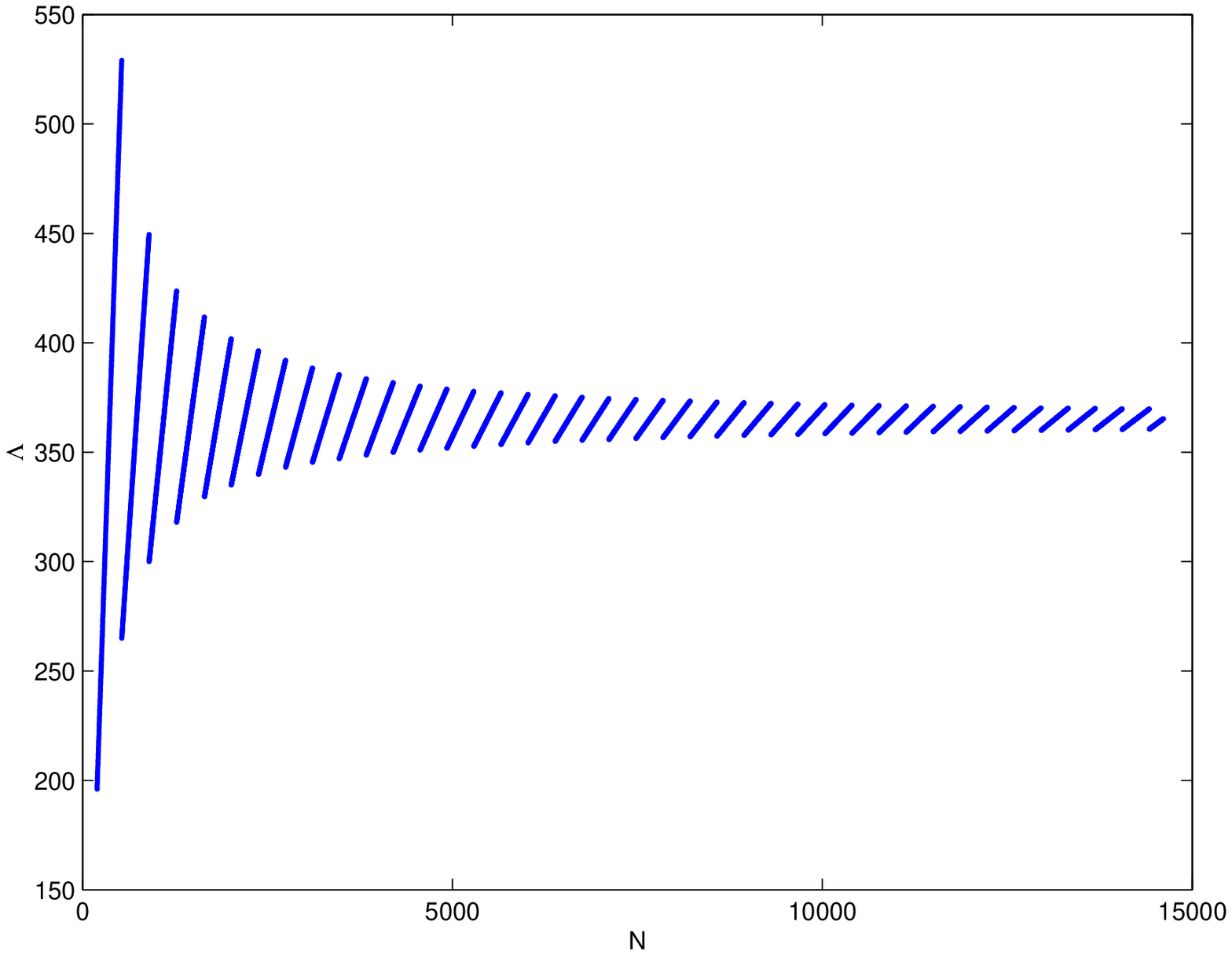}
\caption{The period $\Lambda$ as a function of the number of
points $N$.} \label{fig_om_N} \eec \eef
The uncertainty is dramatic. Even with as many points as $7,920 -
8,140$ (more than $20$ times the `true' period corresponding to
the tropical year), the period takes values in the range $360 -
370$ days. In experimental situations where the value is not
foreseeable from the very beginning, such uncertainty would be
disastrous for estimating statistical properties such as variance
and autocorrelation function. Let us suppose that we are given
only the first $N_1 = 7,920$ data points. We would find a peak at
$k_{\text{max}}=23$, which corresponds to $\Lambda_1 = 360$. If we
had the first $N_2 = 8,140$ data, we would find a period of
$\Lambda_2 = 370$, even with the same $k_{\text{max}}=23$. In
order to fully identify the deterministic component $d_j$ we use
the ansatz \beq d_j = \sum_{q=1}^{Q}{\left[ a_q \cos \frac{2\pi
q}{\Lambda}j + b_q \sin \frac{2\pi q}{\Lambda}j \right]},
\label{eq_dj} \eeq where $Q$ is the number of harmonics used in
the estimation. In each case we minimize the least square distance
between the signal diminished by its mean and the expression in
(\ref{eq_dj}). For $Q=3$, we obtain: \vskip 12pt \bec
\begin{tabular}{|l|ccccccc|}
 \hline
 \  {} && $a_1$ & $b_1$ & $a_2$ & $b_2$ & $a_3$ & $b_3$ \\
  \hline
  $\Lambda_1=360$ && -2.02 & -10.58 & -0.69 & -0.51 & -0.02 & -0.27 \\
  $\Lambda_2=370$ && -9.54 & 5.22 & 1.18 & -0.18 & 0.07 & 0.17 \\
  \hline
\end{tabular}
\eec
\vskip 12pt\noindent
The variances of the residuals are $\sigma_1^2 = 49.17$ and
$\sigma_2^2 = 47.33$ for $\Lambda_1$ and $\Lambda_2$. By using the
tropical year estimate ($\Lambda_{\text{tr}} = 365.2422$), we get
$\sigma_{\text{tr}}^2 = 26.65$ for $N=7,920$, and
$\sigma_{\text{tr}}^2 = 26.51$ for $N=8,140$. It is evident that a
wrong value of the period gives rise to an overestimated variance
of the residuals.
\bef \bec
\includegraphics[scale=0.54]{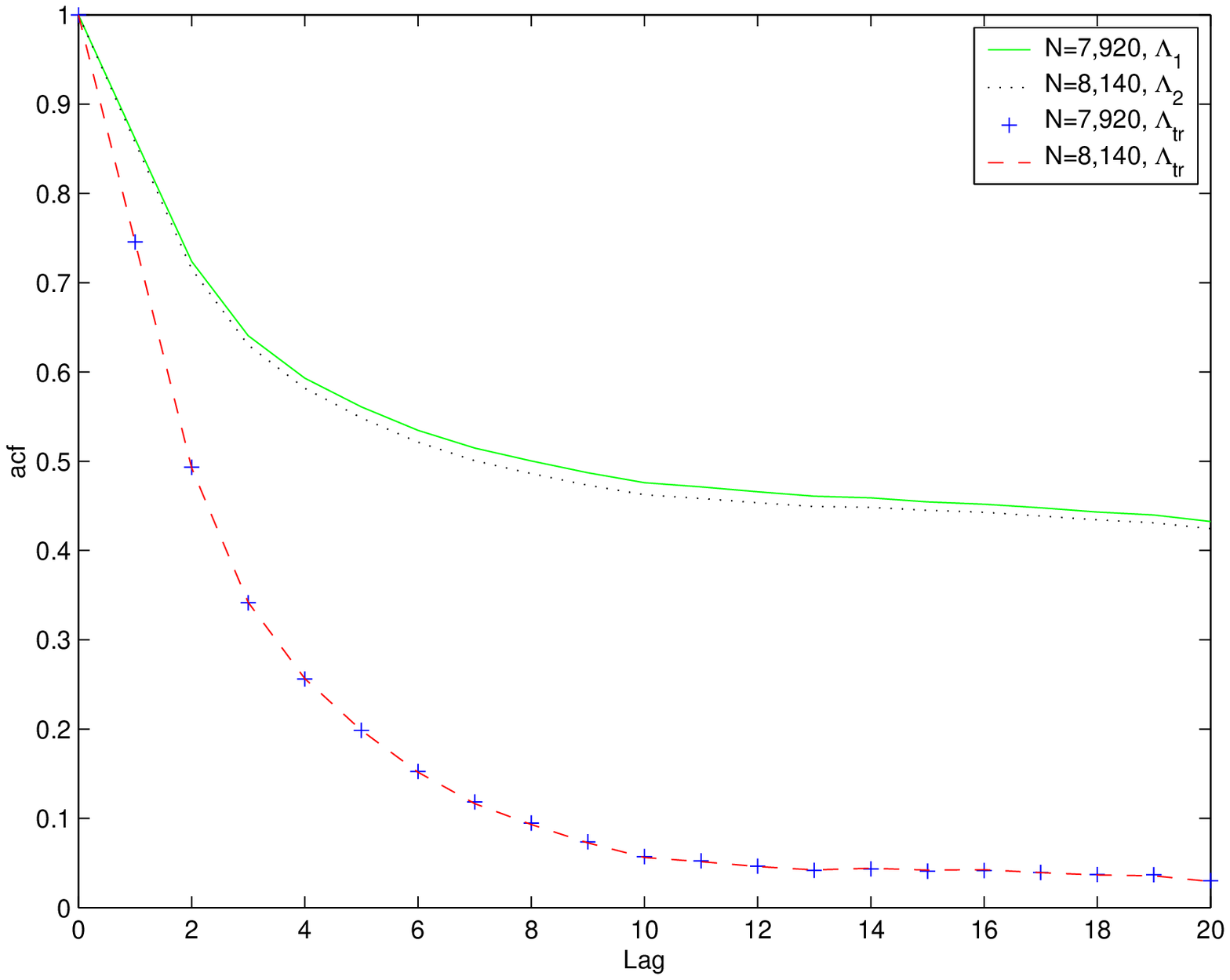}
\caption{Autocorrelation function as a function of the lag.}
\label{fig_acf} \eec \eef
In Figure \ref{fig_acf}, we show how dramatic the effect on the
autocorrelation function is when a wrong period is used. Not only
the residuals have an enlarged variance but they also show a
spurious persistence. In the case of temperature time series, a
large persistence would imply the possibility of forecasting the
weather beyond any reasonable range. \vskip6pt As discussed above,
the determination of the `true' frequency is paramount for an
effective analysis of statistical properties of a signal. Here we
provide a solution to this problem by analyzing the functional
dependence of the peaks appearing in Figure \ref{fig_om_N}. More
than providing the solution, we want to stress that any time the
DFT is performed on a finite sample data set a figure like Figure
\ref{fig_om_N} should be generated. One should not seek the
largest possible data set, but rather study the dependence of the
estimated period as a function of the number of points. Our ansatz
is that both the max and the min `peaks' of each piecewise linear
segment in Figure \ref{fig_om_N} lie on a curve of the type: \beq
\Lambda_{\text{max,min}}(x) = \frac{A x}{B + x}. \label{eq_om_dec}
\eeq By least square minimization, we find: \bec
\begin{tabular}{|c|c|c|}
  \hline
  {} & $A$ & $B$ \\
  \hline
  $\Lambda_{\text{min}}$ & 365.70 & 194.26 \\
  $\Lambda_{\text{max}}$ & 366.19 & -165.67 \\
  \hline
\end{tabular}
\eec In both cases the $R^2$ is $1$ apart from round off errors.
This means that the function in (\ref{eq_om_dec}) is not just a
good guess, but rather the `true' decay of the estimated period to
its large $N$ limit (i.e., $A$). The results in the above table
show that the estimates are fairly close to the `true' period.
Furthermore, the variance and the autocorrelation function are
very close to those computed for the tropical year value,
$365.2422$. To be more specific, the variance for $N=8,140$ is
$26.79$ and $27.54$ for $\Lambda_{\text{min}}$ and
$\Lambda_{\text{max}}$, respectively. In both cases, the
autocorrelation function cannot be distinguished from the one
obtained after using $\Lambda_{\text{tr}}$.

It is worth pointing out that equation (\ref{eq_FT_mnchr_inc}) is
of some help for understanding a more fundamental reason for the
functional dependence in equation (\ref{eq_om_dec}). Adding a
point to the data set is equivalent to changing the lattice or
changing $\eps=\eps(N)$. The jumps appearing in Figure
\ref{fig_om_N} will occur when $\eps$ exceeds $\frac{1}{2}$,
which occurs approximately after adding $\sim \Lambda$ points. By
locating the minima and maxima in this way, one can obtain an
expression for the decay of the form (\ref{eq_om_dec}).

\section{Conclusion}
In conclusion, we have shown how to compute the autocorrelation
function of noisy periodic signals in the case of a single
frequency mode. Our scheme improves upon a naive application of
the DFT. The main point is that more is not better when it comes
to the DFT. The dependence on the number of points for the
estimated frequency is more important than the position of the
peak for a specific $N$. Our solution stems from building
envelopes of minima and maxima of piecewise linear functions.
This can certainly be improved; however there is no question that
another solution has to be found by looking at the results in
Figure \ref{fig_om_N}.

Finally, our approach is applicable to signals with more than one
frequency mode. This will be the subject of a future work.


\end{document}